\documentclass[12pt,preprint]{aastex}

\slugcomment{Submitted to \apjl.}

\shorttitle{Polarization of high-energy pulsar radiation in the
  striped wind model} \shortauthors{P\'etri and Kirk}

\newcommand{\eqb}{\begin{eqnarray}}
\newcommand{\eqe}{\end{eqnarray}}
\begin{document}


\title{Polarization of high-energy pulsar radiation in the striped
  wind model}


\author{J. P\'etri\altaffilmark{1} and J.\ G.\ Kirk\altaffilmark{1}}
\affil{Max-Planck-Institut f\"ur Kernphysik, Saupfercheckweg 1,
  69117 Heidelberg, Germany.}
\email{Jerome.Petri@mpi-hd.mpg.de}
\email{John.Kirk@mpi-hd.mpg.de}



\begin{abstract}
  The Stokes parameters of the pulsed synchrotron radiation produced
  in the striped pulsar wind model are computed and compared with
  optical observations of the Crab pulsar.  We assume the main
  contribution to the wind emissivity comes from a thin transition
  layer where the dominant toroidal magnetic field reverses its
  polarity. The radial component of the field is neglected, but a
  small meridional component is added.  The resulting radiation is
  linearly polarized (Stokes $V=0$). In the off-pulse region, the
  electric vector lies in the direction of the projection on the sky
  of the rotation axis of the pulsar. This property is unique to the
  wind model and in good agreement with the data. Other properties
  such as a reduced degree of polarization and a characteristic sweep
  of the polarization angle within the pulses are also reproduced.
  These properties are qualitatively unaffected by variations of the
  wind Lorentz factor, the electron injection power law index and the
  inclination of the line of sight.
\end{abstract}





\section{Introduction}

The high-energy, pulsed emission from rotating magnetized neutron
stars is usually explained in the framework of either the polar
cap~\citep{sturrock71,rudermansutherland75} or the outer gap
models~\citep{chenghoruderman86,hirotanishibata99}, for a review
see~\citep{harding01}.  Although the existence of such gaps is
plausible \citep{petrietal02}, these models still suffer from the lack
of a self-consistent solution for the pulsar magnetosphere and are
based on the assumption that the magnetic field structure is that of a
(Newtonian) rotating dipole.  Nevertheless, recent observations of the
polarization of the optical pulses from the Crab
\citep{kellner02,kanbachetal03} motivated detailed comparative studies
of the emission from these models \citep{dykshardingrudak04} as well
as a new variant of the outer gap introduced by~\citet{dyksrudak03}
and called the \lq\lq two-pole caustic\rq\rq\ model. This model has a
gap extending from the light-cylinder to the polar caps.  It
reproduces the twin-peak pulse intensity profiles by associating each
peak with a different magnetic pole.  In all of these models, the
radiation is produced within the light cylinder. However, the pulse
profile is determined by the assumed geometry of the magnetic field
and the location of the gaps.  Despite earlier claims
\citep{romaniyadigaroglu95}, it now appears that neither the polar cap
nor the outer gap model are able to fit the optical polarization
properties of the Crab pulsar. The two-pole caustic model, on the
other hand, provides a qualitatively reasonable fit
\citep{dykshardingrudak04}.

An alternative site for the production of pulsed radiation has been
investigated recently \citep{kirkskjaeraasengallant02}, based on the
idea of a striped pulsar wind, originally introduced by
\citet{coroniti90} and \citet{michel94} and elaborated by
\cite{lyubarskykirk01} and \cite{kirkskjaeraasen03}.  Compared to the
inner, outer and slot gaps, much less progress has been made in
understanding the particle acceleration problem in this model. As
proposed by \citet{lyubarsky96}, it is assumed that magnetic energy is
released by reconnection in the thin regions where the toroidal field
reverses its polarity, producing a non thermal electron/positron
population.  But, unlike in Lyubarsky's model, emission from the
striped wind originates outside the light cylinder and relativistic
beaming effects are responsible for the phase coherence of the
synchrotron radiation. A strength of this model is that the geometry
of the magnetic field, which is the key property determining the
polarization properties of the emission, is relatively well-known.
Because relativistic, magnetically dominated winds seem to collimate
only very weakly \citep{bogovalov01,vlahakis04}, the dominant field
component well outside the light-cylinder is toroidal, irrespective of
the geometry of the magnetic field near the stellar surface.

In this Letter, we use an explicit asymptotic solution for the
large-scale field structure related to the oblique split monopole and
valid for the case of an ultrarelativistic plasma \citep{bogovalov99}.
This is combined with a crude model for the emissivity of the striped
wind and of the magnetic field within the dissipating stripes
themselves. We calculate the polarization properties of the
high-energy pulsed emission by generalising a method applied to
gamma-ray burst models by \citet{lyutikovparievblandford03}, and
discuss and compare our results with optical observations of the Crab
pulsar.

\section{Calculation of the Stokes Parameters}

Our magnetic field model is based on the asymptotic solution of
\citet{bogovalov99}, valid for $r\gg r_{\rm L}$, and modified to take
account of a finite width of the current sheet and with a small
additional meridional component in the sheet, which serves to prevent
the magnetic field from becoming identically zero.  In order to
simplify the Lorentz transformations, the small radial component of
the field is neglected.  In spherical polar coordinates
$(r,\theta,\varphi)$ centered on the star and with axis along the
rotation axis, the radial field is small ($B_r \sim B_{\rm L} \,
r_{\rm L}^2/r^2$) and the other components are: \eqb \{ B_\theta, \;
B_\varphi \} & = & B_{\rm L} \, \frac{r_{\rm L}}{r} \, \{ b_{1,2} \,
\eta_{\theta}(\Delta_\theta, r,\theta,\varphi,t), \;
\eta_\varphi(\Delta_\varphi,r,\theta,\varphi,t) \}
\nonumber \\
\eta_\varphi(\Delta_\varphi, r,\theta,\varphi,t) & = &
\tanh\left[\Delta_\varphi\, \left( \cos\theta \, \cos\alpha +
    \sin\theta \, \sin\alpha \, \cos \left\{\varphi - \Omega_* \,
      \left( t - \frac{r}{v} \right) \right\} \right) \right]
\nonumber\\
\eta_\theta(\Delta_\theta, r,\theta,\varphi,t) & = &
\frac{1}{\Delta_\theta} \, \frac{\partial \eta_\varphi(\Delta_\theta,
  r,\theta,\varphi,t)}{\partial \varphi}
\label{magneticfield}
\eqe Here, $B_{\rm L}$ is a fiducial magnetic field strength, $v$ is
the (radial) speed of the wind, $\Omega_*=c/r_{\rm L}$ is the angular
velocity of the pulsar, with $r_{\rm L}$ the radius of the light
cylinder, $\alpha$ is the angle between the magnetic and rotation
axes, $b_{1,2}$ are parameters controlling the magnitude of the
meridional field in the two current sheets present in one wavelength,
and $\Delta_{\theta,\varphi}$ are parameters quantifying the sheet
thickness. The functional form of $B_\varphi$ is motivated by exact
equilibria of the planar relativistic current sheet
\cite[see][]{kirkskjaeraasen03}.  However, in these equilibria the
$B_\theta$ component, which has an important influence on the
polarization sweeps, is arbitrary.  The $B_\theta$ we adopt
corresponds to a small circularly polarized component of the pulsar
wind wave, such as is expected if the sheets are formed by the
migration of particles within the wave, as described qualitatively by
\citet{michel71}.
 
For the particle distribution, we adopt an isotropic electron/positron
distribution given by $N(E,\vec{p},\vec{r},t) = K(\vec{r},t) \,
E^{-p}$ where $K(\vec{r},t)$ is related to the number density of
emitting particles.  The radial motion of the wind imposes an overall
$1/r^2$ dependence on this quantity, which is further modulated
because the energization occurs primarily in the current sheet. The
precise value in each sheet is chosen to fit the observed intensity of
each sub-pulse.  In addition, a small dc component is added, giving
the off-pulse intensity.  For the emissivity, we use the standard
expressions for incoherent synchrotron radiation of ultrarelativistic
particles in the Airy function approximation
\citep{ginzburgsyrovatskii69,melrose71}.  Following
\citet{kirkskjaeraasengallant02}, we assume the emission commences
when the wind crosses the surface $r=r_0\gg r_{\rm L}$.

The calculation of the Stokes parameters as measured in the observer
frame involves simply integrating the emissivity over the wind.
However, this requires special care, because the Lorentz boost from
the rest frame of the emitting plasma involves not only beaming and
Doppler shift, but also a change in the polarization angle due to the
effects of aberration.  \citet{lyutikovparievblandford03} performed
this calculation in the context of a gamma-ray burst model, assuming a
relativistic shell of emitting plasma containing only a toroidal
magnetic field.  However, the relatively simple form of the emissivity
function in that case means that two of the four Stokes parameters
integrate to zero: $U=V=0$, corresponding to linear polarization with
constant position angle. Here we extend this method by adding
a~$B_\theta$ component and allowing for a more general, space and
time dependent  emissivity.

After straightforward but lengthy manipulations involving Lorentz
transformations, we find the Stokes parameters as measured by an
observer at time~$t_{\rm obs}$ are given by the following integrals:
\begin{eqnarray}
  \label{eq:StokesParameters}
  \left\{
    \begin{array}{c}
      I_\omega \\
      Q_\omega \\
      U_\omega
    \end{array}
  \right\} (t_{\rm obs}) & = & \int_{r_0}^{+\infty} \int_0^{\pi} \int_0^{2\,\pi}
  s_0(r,\theta,\varphi,t_{\rm ret}) \,  
  \left\{
    \begin{array}{c}
      \frac{p+7/3}{p+1} \\
      \cos \, (2\,\tilde{\chi}) \\
      \sin \, (2\,\tilde{\chi})
    \end{array}
  \right\}
  \, r^2 \, \sin\theta \, dr \, d\theta \, d\varphi
\end{eqnarray}
where the retarded time is given by $t_{\rm ret} = t_{\rm obs} +
\vec{n}\cdot\vec{r}/c$ and $\vec{n}$ is a unit vector along the line
of sight from the pulsar to the observer.  In this approximation the
circular polarization vanishes: $V=0$. The function~$s_0$ is defined
by:
\begin{eqnarray}
  \label{eq:ParaStokes}
  s_0(r,\theta,\varphi,t) & = & \kappa \, K(\vec{r},t) \,
  \omega^{-\frac{p-1}{2}} \, \mathcal{D}^{\frac{p+3}{2}}
  \, \left( \frac{B}{\Gamma} \, \sqrt{ 1 - ( \mathcal{D} \, \vec{n} \cdot \vec{b} )^2 } 
  \right)^{\frac{p+1}{2}}
\end{eqnarray}
where $\omega$ is the (angular) frequency of the emitted radiation, and
$\kappa$ is a constant factor that depends only on the nature of the
radiating particles~(charge $q$ and mass $m$) and the power law index
$p$ of their distribution:
\begin{eqnarray}
  \kappa & = & \frac{\sqrt{3}}{2\,\pi} \, \frac{1}{4} \, \Gamma_{\rm Eu} 
  \left( \frac{3\,p+7}{12}\right) \, 
  \Gamma_{\rm Eu}\left(\frac{3\,p-1}{12}\right) \, 
  \frac{|q|^3}{4\,\pi\,\varepsilon_0\,m\,c} 
  \, \left( \frac{3\,|q|}{m^3\,c^4} \right)^{\frac{p-1}{2}} 
\end{eqnarray}
with $\Gamma_{\rm Eu}$ the Euler gamma function and $\mathcal{D}$ the
Doppler boosting factor $\mathcal{D} = 1/\Gamma \, ( 1 - \vec{\beta}
\cdot \vec{n} )$. The direction of the local magnetic field in the
observer's frame is given by the unit vector $\vec{b}$ and the
simplifying assumption has been made that this field has no component
in the direction of the plasma velocity: $\vec{b}\cdot\vec\beta=0$.
(In this case the magnetic field transformation from the rest
frame~$\vec{B}'$ to the observer frame~$\vec{B}$ is just $\vec{B}' =
\vec{B}/\Gamma$ and, thus, its direction remains unchanged.)  The
angle~$\tilde{\chi}$ measures the inclination of the {\it local\/}
electric field with respect to the projection of the pulsar's rotation
axis on the plane of the sky as seen in the observer's frame. The
degree of linear polarization is defined by $\Pi = \sqrt{Q^2 +
  U^2}/I$. The corresponding polarization angle, defined as the
position angle between the electric field vector at the observer and
the projection of the pulsar's rotation axis on the plane of the sky
is $\chi = 1/2 \, \arctan ( U/Q ) $.

\section{Results}

Using a model similar to that described above,
\citet{kirkskjaeraasengallant02} computed the total radiation
intensity and compared the spacing of the resulting twin-pulse profile
with observations of the Crab pulsar. They found an obliquity
$\alpha=60\degr$ and an inclination of the rotation axis to the line
of sight $\xi=\arccos(\vec{n}\cdot
\vec{\Omega}_*/|\vec{\Omega}_*|)=60\degr$. In the following, we adopt
these parameters, and set the radius at which emission switches on to
be $r_0=30r_{\rm L}$.

\begin{figure}
 \epsscale{.9} \plotone{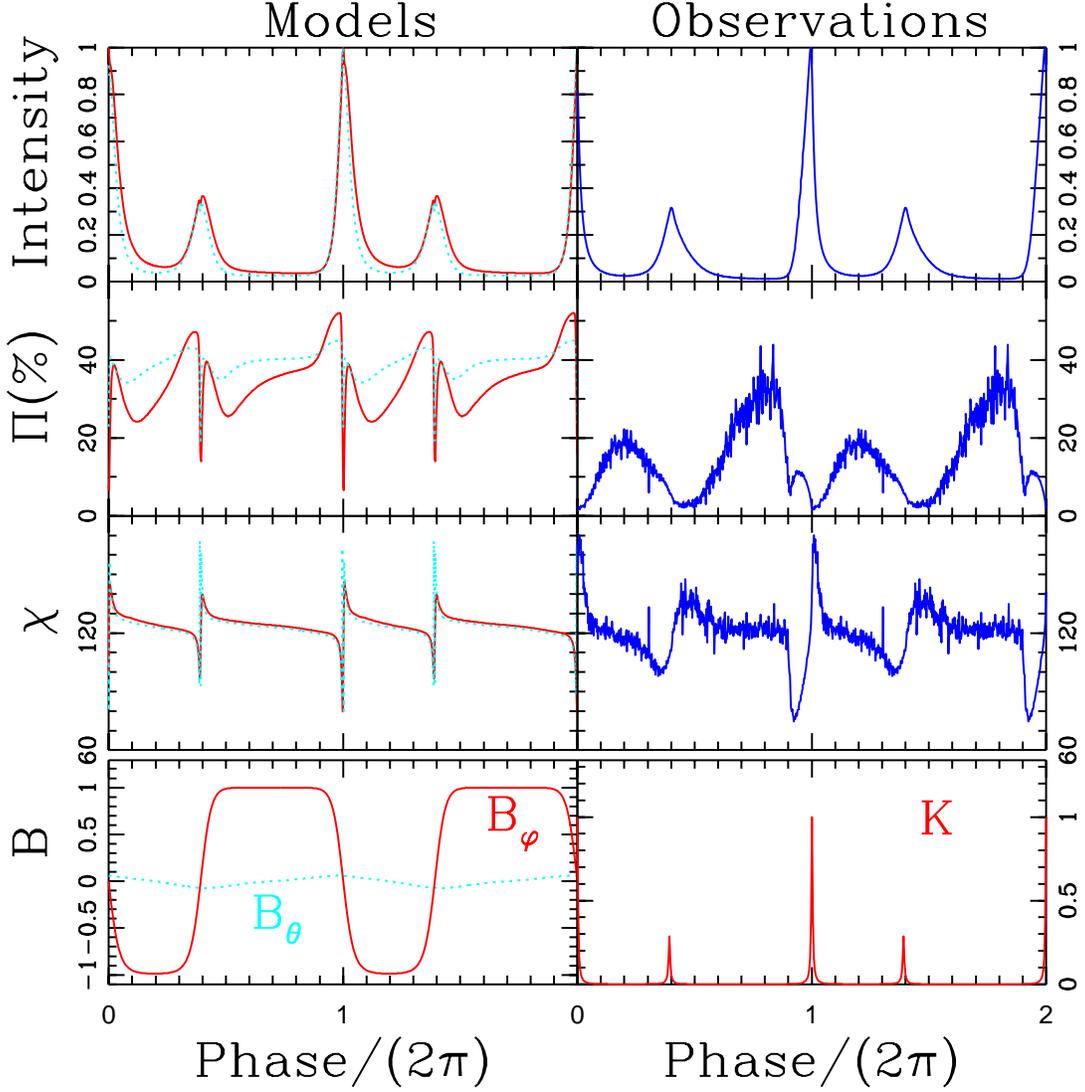}
\caption{Light curve of intensity, degree of polarization and 
  position angle of the pulsed synchrotron emission obtained by
  numerical integration of the set of
  equations~(\ref{eq:StokesParameters}), and measurements of these
  quantities for the Crab pulsar \citep{kellner02,kanbachetal03}.
  Models with Lorentz factor~$\Gamma=20$ (solid line, red online) and
  $50$ (dotted line, cyan online) are shown.  The assumed particle
  energy distribution index was $p=2$, and the inclination of the line
  of sight equals the obliquity: $\alpha = \xi = 60\degr$. The
  position angle of the projection of the pulsar's rotation axis was
  set to $124\degr$ \citep{ngromani04}.  The bottom panels show the
  dependence on phase ($=\Omega_*r/(2\pi v)$ in
  Eq.~(\ref{magneticfield})) of the assumed magnetic field components
  and the particle density in the comoving frame.  The maximum values
  of $B_\varphi$ and $K$ are normalized to unity.
  \label{fig1}}
\end{figure}

In the model of \citet{kirkskjaeraasengallant02}, the current sheet
was assumed to be thin, which results in sharp profiles with very
similar shapes for each of the subpulses. The upper left panel of
Fig.~\ref{fig1} shows the intensity (Stokes parameter $I$) computed
using our smoothed profile with $\Delta_\theta=1$, $\Delta_\varphi=5$,
$b_1=0.1$ and $b_2=0.08$ for each subpulse. The electron density is $K
= [ \{ ( r_{\rm L} \, B_{\rm L} ) / ( r \, B ) \}^{(p+1)/2} +
\varepsilon - 1 ] / [ r^2 \, ( 1 - 0.6\,\eta_\theta) ]$ where the
parameter $\varepsilon=0.05$ sets the minimum electron density between
the current sheets (in normalized units). The
denominator~$(1-0.6\,\eta_\theta)$ introduces an asymmetry in the
relative pulse peak intensity. The variation of the magnetic field and
the particle density along the line of sight, are shown in the bottom
panels of Fig.~\ref{fig1}.

The upper panels of Fig.~\ref{fig1} show the results of our
computations on the left and the corresponding observed quantities
\citep{kellner02,kanbachetal03} on the right.  It should be noted that
these data are preliminary. In particular, the measured degree of
polarization may be subject to revision (Kanbach priv.\ comm.).
Comparison with the upper right-hand panel shows that the model
reproduces the observed pulse profile quite accurately.  However, the
idealised transverse geometry of the magnetic field in the (presumably
turbulent) sheet leads to a rapid variation in phase of the term in
parentheses on the right-hand side of Eq.~(\ref{eq:ParaStokes}),
giving rise to a small notch-like feature visible in the peak of the
sub-pulse, that proves difficult to eliminate.  In this example, we
adopted an electron power law index $p=2$, as suggested by the
relatively flat spectrum displayed by the pulsed emission between
optical and gamma-ray frequencies
\citep{shearergolden01,kanbach98,kuiperetal01}.  Results are shown for
two values of the Lorentz factor of the wind: $\Gamma=20$ (solid line)
and $\Gamma=50$ (dotted line).  For convenience, the maximum intensity
is normalized to unity. The timescale is expressed in terms of pulse
phase, $0$ corresponding to the initial time~$t=0$ and $1$ to a full
revolution of the neutron star and thus one period~$t = 2\pi /
\Omega_*$.

The degree of polarization is shown in the two middle panels of
Fig.~\ref{fig1}. According to our computations (left-hand panel) this
displays a steady rise in the initial off-pulse phase, that steepens
rapidly as the first pulse arrives.  During the pulse phase itself,
the polarization shrinks down to about 10\%. Theoretically the maximum
possible degree of polarization is closely related to the index~$p$ of
the particle spectrum.  In the most favorable case of a uniform
magnetic field, it is given by $\Pi_{\rm max} = (p+1)/(p+7/3)$.

However, in the curved magnetic field lines of the wind, contributions
of electrons from different regions have different polarization
angles. Consequently, they depolarize the overall result when
superposed. We therefore expect a degree of polarization that is at
most~$\Pi_{\rm max}$. For the example shown in figure~\ref{fig1},
$\Pi_{\rm max}(p=2) = 69.2$\%, well above the computed value, which
peaks at~$52$\%.

The lower panels of Fig.~\ref{fig1} show the polarization angle,
measured against celestial North and increasing from North to East.
Our model predicts this angle relative to the projection of the
rotation axis of the neutron star on the sky, which we take to lie at
a position angle of $124\degr$, following the analysis of
\citet{ngromani04}.  In the off-pulse stage, the electric vector of
our model predictions lies almost exactly in this direction, since it
is fixed by the orientation of the dominant toroidal
component~$B_\varphi$ of the magnetic field.  Such a relation between
the off-pulse angle of polarization and the toroidal magnetic field at
large distance ({\bf close to} the light cylinder) was already
suggested by \cite{smithetal88}.  In the rising phase of the first
pulse, $B_\varphi$ decreases, whereas $B_\theta$ increases, causing
the polarization angle to rotate from its off pulse value by about
$50\degr$, for the chosen parameters.  However, for a weaker
$B_\theta$ contribution, as in the second pulse, the swing decreases.
This effect can also be caused by a relatively large beaming angle,
(i.e., low Lorentz factor wind).  The basic reason is that
contributions from particles well away from the sheet center are then
mixed into the pulse, partially canceling the contribution of the
particles in the center of the current sheet, which favor
$\chi=124\degr\pm90\degr$, and enforcing~$\chi=124\degr$.  On the
other hand, a very high value of the Lorentz factor or large values of
$b_{1,2}$ reduce the off-center contribution, leading, ultimately, to
the maximum possible $90\degr$ sweep between off-pulse
($B_\varphi$-dominated) and center-pulse ($B_\theta$-dominated)
polarizations, followed by another $90\degr$ sweep in the same sense
when returning to the off-pulse. Thus, in general, in the middle of
each pulse, the polarization angle is either nearly parallel to the
projection of the rotation axis, or nearly perpendicular to it,
depending on the strength of~$B_\theta$ and on~$\Gamma$.  This
interpretation is confirmed by computations with $\Gamma=50$ that show
a larger sweep, as shown in Fig.~\ref{fig1}.

The optical polarization measurements suggest that in the centre of
the pulses the position angle is close to $124\degr$.  In the
declining phase of both pulses, the angle reaches a maximum before
returning to the off-pulse orientation.  Note that in both cases the
swing starts in the same direction, (counterclockwise in
figure~\ref{fig1}). This is determined by the rotational behavior of
the $B_\theta$ component, implying that this changes sign between
adjacent sheets, as in Eq.~(\ref{magneticfield}).  The observed
off-pulse position angle is closely aligned with the projection of the
rotation axis of the pulsar, in accordance with the model predictions.

In addition to models aimed at providing a framework for the
interpretation of the emission of the Crab pulsar, we have performed
several calculations with different Lorentz factors~$\Gamma$,
injection spectrum of relativistic electrons~$p$ and inclinations of
the line of sight~$\xi$. The general characteristics of the results
are: For low Lorentz factors, independent of $p$ and $\xi$, the
relativistic beaming becomes weaker and the pulsed emission is less
pronounced, because the observer receives radiation from almost the
entire wind. For instance, taking~$\Gamma=2$ and $p=2$ or $3$, the
average degree of polarization does not exceed 20~\% and the swing in
the polarization angle is less than $30\degr$. For high Lorentz
factors $\Gamma\ge50$, the strong beaming effect means that the
observer sees only a small conical fraction of the wind. The width of
the pulses is then closely related to the thickness of the transition
layer. The degree of linear polarization flattens in the off-pulse
emission while it shows a sharp increase followed by a steep decrease
during the pulses. Due to the very strong beaming effect, only a tiny
part of the wind directed along the line of sight will radiate towards
the observer. In the off-pulse phase, the polarization angle is then
dictated solely by the $B_\varphi$ component \lq\lq attached\rq\rq\ to
the line of sight, and the degree of linear polarization remains
almost constant in time.  For very high Lorentz factors, the behavior
of polarization angle and degree remain similar to those of
Fig.~\ref{fig1}, with perfect alignment between polarization direction
(electric vector) and the projection of the pulsar's rotation axis on
the plane of the sky in the off-pulse phase and two consecutive
polarization angle sweeps of $90\degr$ in the same sense during the
off-pulse to center-pulse and center-pulse to off-pulse transitions.
This mirrors the fact that emission comes only from a narrow cone
about the line of sight of half opening angle~$\theta \approx 1 /
\Gamma$.

For given values of~$\Gamma$ and $\xi$, the particle spectral index
$p$ affects only the average degree of polarization degree but not the
light curve nor the polarization angle. For example, taking
$\Gamma=10$ and~$\xi=60\degr$, a spectral index of~$p=2$ leads to
an average polarization of~$\tilde{\Pi}=19.2\%$ whereas for~$p=3$ it
leads to~$\tilde{\Pi}=30.8\%$.

\section{Conclusions}

In the striped wind model, the high energy (infra-red to gamma-ray)
emission of pulsars arises from outside the light cylinder, in
accordance with the early suggestions of \citet{pacinirees70} and
\citet{shklovsky70}.  It provides an alternative to the more
intensively studied polar cap, outer gap and two-pole caustic models.
All models contain essentially arbitrary assumptions concerning the
configuration of the emission region and the distribution function of
the emitting particles, rendering it difficult to distinguish between
them on the basis of observations. However, the geometry of the
magnetic field, which is the crucial factor determining the
polarization properties, is constrained in the striped model to be
close to that of the analytic asymptotic solution presented by
\cite{bogovalov99}.  We have therefore presented detailed computations
of the polarization properties of the pulses expected in this
scenario. These possess the characteristic property, unique amongst
currently discussed models, that the electric vector of the off-pulse
emission is aligned with the projection of the pulsar's rotation axis
on the plane of the sky.  This is in striking agreement with recent
observations of the Crab pulsar.  In addition the striped wind
scenario naturally incorporates features of the phase-dependent
properties of the polarization angle, degree of polarization and
intensity that are also seen in the data.  This underlines the need to
develop the model further, in order to confront high-energy
observations of the Crab and other pulsars.  In particular, the manner
in which magnetic energy is released into particles in the current
sheet remains poorly understood and the link between the asymptotic
magnetic field structure and the pulsar magnetosphere is obscure.

\acknowledgements
We thank Gottfried Kanbach for providing us with the OPTIMA data and for
helpful discussions. This work was supported by a grant from the 
G.I.F., the German-Israeli Foundation for Scientific Research and
Development. 




\begin{thebibliography}{31}
\expandafter\ifx\csname natexlab\endcsname\relax\def\natexlab#1{#1}\fi

\bibitem[{{Bogovalov}(1999)}]{bogovalov99}
{Bogovalov}, S.~V. 1999, \aap, 349, 1017

\bibitem[{{Bogovalov}(2001)}]{bogovalov01}
---. 2001, \aap, 371, 1155

\bibitem[{{Cheng} {et~al.}(1986){Cheng}, {Ho}, \&
  {Ruderman}}]{chenghoruderman86}
{Cheng}, K.~S., {Ho}, C., \& {Ruderman}, M. 1986, \apj, 300, 500

\bibitem[{{Coroniti}(1990)}]{coroniti90}
{Coroniti}, F.~V. 1990, \apj, 349, 538

\bibitem[{{Dyks} {et~al.}(2004){Dyks}, {Harding}, \&
  {Rudak}}]{dykshardingrudak04}
{Dyks}, J., {Harding}, A.~K., \& {Rudak}, B. 2004, \apj, 606, 1125

\bibitem[{{Dyks} \& {Rudak}(2003)}]{dyksrudak03}
{Dyks}, J., \& {Rudak}, B. 2003, \apj, 598, 1201

\bibitem[{{Ginzburg} \& {Syrovatskii}(1969)}]{ginzburgsyrovatskii69}
{Ginzburg}, V.~L., \& {Syrovatskii}, S.~I. 1969, \araa, 7, 375

\bibitem[{{Harding}(2001)}]{harding01}
{Harding}, A.~K. 2001, in American Institute of Physics Conference Series
  \#558, 115

\bibitem[{{Hirotani} \& {Shibata}(1999)}]{hirotanishibata99}
{Hirotani}, K., \& {Shibata}, S. 1999, \mnras, 308, 54

\bibitem[{{Kanbach}(1998)}]{kanbach98}
{Kanbach}, G. 1998, Advances in Space Research, 21, 227

\bibitem[{{Kanbach} {et~al.}(2003){Kanbach}, {Kellner}, {Schrey}, {Steinle},
  {Straubmeier}, \& {Spruit}}]{kanbachetal03}
{Kanbach}, G., {Kellner}, S., {Schrey}, F.~Z., {Steinle}, H., {Straubmeier},
  C., \& {Spruit}, H.~C. 2003, in Instrument Design and Performance for
  Optical/Infrared Ground-based Telescopes. Edited by Iye, Masanori; Moorwood,
  Alan F. M. Proceedings of the SPIE, Volume 4841, pp. 82-93 (2003)., 82--93

\bibitem[{{Kellner}(2002)}]{kellner02}
{Kellner}, S. 2002, Master's thesis, Technische Universit\"at M\"unchen

\bibitem[{{Kirk} \& {Skj{\ae}raasen}(2003)}]{kirkskjaeraasen03}
{Kirk}, J.~G., \& {Skj{\ae}raasen}, O. 2003, \apj, 591, 366

\bibitem[{{Kirk} {et~al.}(2002){Kirk}, {Skj{\ae}raasen}, \&
  {Gallant}}]{kirkskjaeraasengallant02}
{Kirk}, J.~G., {Skj{\ae}raasen}, O., \& {Gallant}, Y.~A. 2002, \aap, 388, L29

\bibitem[{{Kuiper} {et~al.}(2001){Kuiper}, {Hermsen}, {Cusumano}, {Diehl},
  {Sch{\" o}nfelder}, {Strong}, {Bennett}, \& {McConnell}}]{kuiperetal01}
{Kuiper}, L., {Hermsen}, W., {Cusumano}, G., {Diehl}, R., {Sch{\" o}nfelder},
  V., {Strong}, A., {Bennett}, K., \& {McConnell}, M.~L. 2001, \aap, 378, 918

\bibitem[{{Lyubarskii}(1996)}]{lyubarsky96}
{Lyubarskii}, Y.~E. 1996, \aap, 311, 172

\bibitem[{{Lyubarsky} \& {Kirk}(2001)}]{lyubarskykirk01}
{Lyubarsky}, Y., \& {Kirk}, J.~G. 2001, \apj, 547, 437

\bibitem[{{Lyutikov} {et~al.}(2003){Lyutikov}, {Pariev}, \&
  {Blandford}}]{lyutikovparievblandford03}
{Lyutikov}, M., {Pariev}, V.~I., \& {Blandford}, R.~D. 2003, \apj, 597, 998

\bibitem[{{Melrose}(1971)}]{melrose71}
{Melrose}, D.~B. 1971, \apss, 12, 172

\bibitem[{{Michel}(1971)}]{michel71}
{Michel}, F.~C. 1971, Comments on Astrophysics and Space Physics, 3, 80

\bibitem[{{Michel}(1994)}]{michel94}
---. 1994, \apj, 431, 397

\bibitem[{{Ng} \& {Romani}(2004)}]{ngromani04}
{Ng}, C.-Y., \& {Romani}, R.~W. 2004, \apj, 601, 479

\bibitem[{{P{\' e}tri} {et~al.}(2002){P{\' e}tri}, {Heyvaerts}, \&
  {Bonazzola}}]{petrietal02}
{P{\' e}tri}, J., {Heyvaerts}, J., \& {Bonazzola}, S. 2002, \aap, 384, 414

\bibitem[{{Pacini} \& {Rees}(1970)}]{pacinirees70}
{Pacini}, F., \& {Rees}, M.~J. 1970, \nat, 226, 622

\bibitem[{{Romani} \& {Yadigaroglu}(1995)}]{romaniyadigaroglu95}
{Romani}, R.~W., \& {Yadigaroglu}, I.-A. 1995, \apj, 438, 314

\bibitem[{{Ruderman} \& {Sutherland}(1975)}]{rudermansutherland75}
{Ruderman}, M.~A., \& {Sutherland}, P.~G. 1975, \apj, 196, 51

\bibitem[{{Shearer} \& {Golden}(2001)}]{shearergolden01}
{Shearer}, A., \& {Golden}, A. 2001, \apj, 547, 967

\bibitem[{{Shklovsky}(1970)}]{shklovsky70}
{Shklovsky}, I.~S. 1970, \apjl, 159, L77

\bibitem[{{Smith} {et~al.}(1988){Smith}, {Jones}, {Dick}, \&
  {Pike}}]{smithetal88}
{Smith}, F.~G., {Jones}, D.~H.~P., {Dick}, J.~S.~B., \& {Pike}, C.~D. 1988,
  \mnras, 233, 305

\bibitem[{{Sturrock}(1971)}]{sturrock71}
{Sturrock}, P.~A. 1971, \apj, 164, 529

\bibitem[{{Vlahakis}(2004)}]{vlahakis04}
{Vlahakis}, N. 2004, \apj, 600, 324

\end{thebibliography}

\end{document}